# Gravitational Gauss' Law in the Laboratory: A 21 century Archimedes problem

Timir Datta

Physics & Astronomy

University of South Carolina

Columbia, SC 29208

datta@sc.edu

Key ideas: Density determination from a distance, gravitational imaging, gravitational flux and gravitational Gauss' law.

Gravitational field of many common objects are small but in the detectable μGal-nGal (g, mean free fall acceleration on earth ~10 m/s$^2$) range. Here we show that by applying gauss' law to the exterior field data one can estimate the interior mass density of the object. With the development of wide band width, nano ($10^{-9}$/s$^2$) or Eotovs range accelerometers it may also be possible to reconstruct and image the distribution of interior matter distribution.

## Gravitation the first fundamental force of nature

Gravitation is one of the four fundamental "interactions" of nature. It is characterized by Newton's universal constant of gravitation, G or by the Planck mass, $M_p = (\hbar c/G)^{1/2} \sim 10^{-8}$ kg where, c is the speed of light, ℏ is Planck constant [1 & 2]. In spite of being the "first physical law" and its four century long history, the full theoretical understanding gravitation remains elusive. Also, because of the small value of G (6.67x10$^{-11}$ SI), gravitational forces of laboratory objects are rather weak. Perhaps not coincidentally except for the critical general relativistic corrections in global positioning systems (GPS) and some geophysical prospecting, gravity has not witnessed the explosion of technological applications associated with electromagnetism and quantum mechanical systems.

We reason that with present day gravitational sensors the mass and density of many manmade objects can be determined from a distance. Furthermore with the development of faster and higher resolution accelerometers gravitational tomography may also become a reality. In this context we are restricting this discussion to conventional understanding of gravitation in 3+1 space time. Also not referring to the astrophysical phenomenon of "gravitational lensing and imaging", instead we imply visualization techniques such as scanning probe microscopy to X-ray CAT scans.

To estimate the scale of the gravitational field (**F**) we are dealing with let us calculate the (horizontal) acceleration close to the surface of a concrete column of radius (R) one meter, using cylindrical symmetry and a concrete density of $2.3 \times 10^3$ kg/m$^3$ we obtain:

$$F = \frac{2G(M/h)}{R} = \frac{2G\lambda}{R}$$
$$\sim 10^{-6} \, m/s^2 \quad (1)$$
$$\sim 0.1 \, \mu Gal$$

In equation 1, $\lambda$ is the linear density of the column material. Note to calculate $\lambda$ from equation 1 it is not enough just to measure acceleration in the µGal regime. Because at any one point **F** is the vector superposition of all the fields produced by the gravitating matter distributed everywhere in the universe! In addition, the extraneous gravity contributions are often time dependant [3] and such temporal variations can be significant.

Gravitational gauss' law

One way to avoid the dc-background is by gauss' law which relates the total gravitational flux $\Phi_{grav}$ to M, the total mass enclosed as follows

$$\Phi_{grav} = \iint_{surface} F \bullet ds = -4\pi GM \qquad (2)$$

For equation 2 to be useful, the surface has to enclose the source. Furthermore, in the case of cylindrical symmetry and small diameter to length ratio, or close to the cylindrical surface one obtains:

$$\Phi_{grav} / \Delta z \equiv \phi = -4\pi G \lambda_{ave} \qquad (3)$$

Where, $\phi$ is the average flux per unit height $\Delta z$. It can also be shown that with cylindrical symmetry the integral of equation 2 can be approximated by a finite sum of field data (**F**) obtained from N measurements at vector positions **R**(x,y) as follows,

$$\phi = -2\pi \frac{\sum_{1}^{N}(xF_x + yF_y)_i}{N} = -2\pi \bullet Sum_{ave} \qquad (4)$$

From eq. 3 and eq. 4 we obtain,

$$\lambda_{ave} = \frac{1}{2} Sum_{ave} \qquad (5)$$

As shown in the equations above the total flux is related to the surface integral of the symmetric or interior products of **R** and **F**, it can also be shown that the asymmetric or exterior product **R**x **F** is related to the net torque $\tau$. Net torque will be zero for central fields.

A simple example

To test these ideas let us consider a synthetic problem of three vertical cylinders of mass density $\lambda_1$, $\lambda_2$ and $\lambda_3$ be located at positions **r**$_1$, **r**$_2$, and **r**$_3$ as indicated in figure 1.

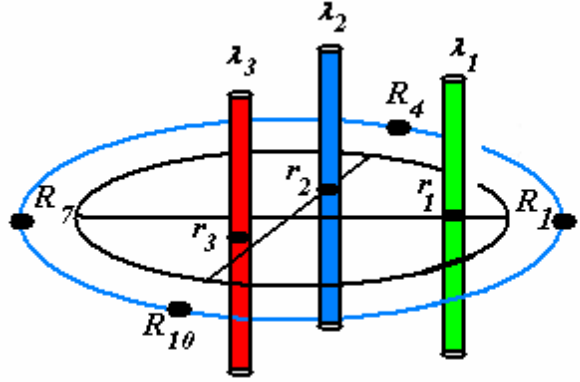

**Figure-1:** A mock up distribution of three cylindrical source masses of linear density $\lambda_i$ located at positions $r_i$ within a circular boundary (black) are shown. The fields **F** due to the cylinders, at twelve points $R_i$ on an exterior (blue) circle of radius R are considered. For clarity only four positions $R_1$, $R_4$, $R_7$ & $R_4$ are marked.

This is a case of a non uniform mass distribution of total linear mass density of 6 $\lambda_1$. For this synthetic demonstration let us first calculate the Newtonian field **F**(R) at the measurement or field points (**R**) due to the three cylinders as given by:

$$F(R) = k\Sigma_1^3 \lambda_i \frac{(r_i - R)}{(r_i - R)^2} \qquad (6)$$

The constant k, in equation 6 contains all the dimensional factors such as G and the mass and length scales of the problem. To be specific, let us compute the **F** fields (as in fig. 1) at just twelve points $R_i$ ($1 \leq i \leq 12$) at 30 deg intervals on the circumference of the exterior circle of radius R equal to 10 units. The mass densities be $\lambda_1=1$, $\lambda_2=2*\lambda_1$ and $\lambda_3 =3*\lambda_1$ respectively located at positions $r_1$ (9,0) $r_2$ (0,5) and $r_3$ (-3,-1). Table 1 gives the calculated values of the various quantities pertaining to the equations above. Columns one and two are the (polar) coordinates of the measurement points, the third column provides the field (in relative units) data and four is polar angle of field direction. Five and six give the contribution to the flux and torque sums respectively.

The changes in both the magnitude and direction of the field as listed in table 1 are indications of a non symmetric density (mass) distribution. To help visualizing these results especially the inhomogeneous distribution a three-d map of the field is displayed in fig.2.

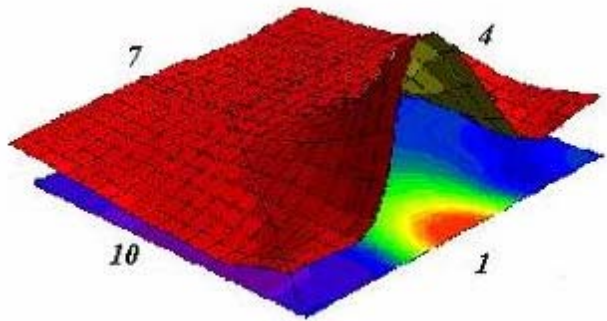

**Figure 2:** A surface plot and contour map of the synthetic field data.

Remarkably, a dozen exterior measurements are adequate to provide a reasonable estimate of the enclosed (mass) density. Quantitatively, the field changes by ~20%, the average of the flux $\phi$ is 6.2, within a few per cent of the exact value of 6 units and the total torque is found to be zero. As mentioned above, a casual perusal of the field map shown in figure 2 makes the non-uniformity of the interior mass distribution eminently clear. It appears that with present day sensors and computers it may be possible to tell if there are large air pockets or perhaps even determine the locations of the steel reinforcement bars inside a concrete pillar. Incidentally, here we are considering the entire gravitational field at the field point and not doing gradiometry; gravitational gradiometry is relevant in geophysical prospecting.

## A topical iteration of a 3[rd] century BC conundrum

To the best of present day knowledge of all the fundamental interactions gravity cannot be screened, obscured or controlled [4]. Can this unique property be useful? First, recall that in the 3[rd] century BC, Hiero, the king of Syracuse is reputed to have inquired Archimedes about the purity of a gold crown; in modern terms the physical question was how to find the volume of an irregularly shaped object, the crown. From volume and total mass, calculating the average density, the distinguishing parameter of gold purity was reduced to a trivial matter. Now imagine a 21 century, say a plutonium ingot is concealed inside a block of lead and that by a cleaver design of void spaces the overall density and center of mass are matched to bulk lead. The conundrum is how to reveal the dense embedded object by exterior measurements alone. No matter how perfect the radiation shielding of lead may be the gravitational signature of the anomalous regions will show up! Furthermore, from the gravitational field it may be possible to reconstruct the hidden mass distribution.

## Concluding remarks

Because gravitational field of human size objects is weak the idea of imaging with gravitaty may appear "not to work in principle" – but a lot of things have been proven to work when the appropriate principles are recognized [5]. Aside from some differences such as: (i) no external source of radiation. (ii) Source and detector are not both under control only the detector in control and (iii) gravitational influence is not restricted to be along line of sight, the concept presented here is really very similar to CAT scan. With the advent of faster and more sensitive accelerometers such imaging processes may become practical.

Table-1

| R(R.U) | $R_\theta$ (deg) | F (R.U) | $F_\theta$ (deg) | Flux $\phi_i$ | Torque $\tau_i$ |
|---|---|---|---|---|---|
| 10 | 0 | 60.4 | 173.35 | -6 | 70 |
| 10 | 30 | 56.8 | 203.86 | -5.7 | 60.6 |
| 10 | 60 | 54.1 | 236.3 | -5.4 | 35 |
| 10 | 90 | 53 | -90 | -5.3 | 0 |
| 10 | 120 | 54.1 | -56.3 | -5.4 | -35 |
| 10 | 150 | 56.8 | -23.86 | -5.7 | -60.6 |
| 10 | 180 | 60.4 | 6.65 | -6 | -70 |
| 10 | 210 | 63.8 | 35.43 | -6.4 | -60.6 |
| 10 | 240 | 66.2 | 63.05 | -6.6 | -35 |
| 10 | 270 | 67 | 90 | -6.7 | 0 |
| 10 | 300 | 66.2 | 116.95 | -6.6 | 35 |
| 10 | 330 | 63.8 | 144.57 | -6.4 | 60.6 |